\documentclass[aps,prl,showpacs,superscriptaddress,twocolumn,floatfix]{revtex4}
\usepackage{times}
\usepackage{graphicx,color}
\usepackage{amsmath,amssymb}
\def\beq{\begin{equation}}
\def\eeq{\end{equation}}
\def\bea{\begin{eqnarray}}
\def\eea{\end{eqnarray}}

\newcommand{\vect}[1]{\boldsymbol{#1}}
\newcommand{\rme}{\mathrm{e}} %
\newcommand{\rmd}{\mathrm{d}} %
\newcommand{\Wi} {\mathrm{Wi}} %

\newcommand{\eref}[1]{Eq.~(\ref{#1})}%
\newcommand{\Eref}[1]{Equation~(\ref{#1})}%

\begin{document}

\title{Accurate statistics of a flexible polymer chain in shear flow.}

\author{Dibyendu Das}

\affiliation{Department of Physics, Indian Institute of Technology, Bombay,
  Powai, Mumbai-400 076, India}

\author{Sanjib Sabhapandit}
         
\affiliation{Raman Research Institute, Bangalore 560080, India}

\date{\today}

\begin{abstract}
  We present exact and analytically accurate results for the problem
  of a flexible polymer chain in shear flow. Under such a flow the
  polymer tumbles, and the probability distribution of the tumbling
  times $\tau$ of the polymer decays exponentially as $\sim
  \exp(-\alpha \tau/\tau_0)$ (where $\tau_0$ is the longest relaxation
  time). We show that for a Rouse chain, this nontrivial constant
  $\alpha$ can be calculated in the limit of large Weissenberg number
  (high shear rate) and is in excellent agreement with our simulation
  result of $\alpha \simeq 0.324$. We also derive exactly the
  distribution functions for the length and the orientational angles
  of the end-to-end vector $\vect{R}$ of the polymer.
\end{abstract}


\maketitle

Dynamics of a polymer under a shear flow has been of great interest both
experimentally and
theoretically~\cite{gennes,chu,LeDuc,smith,doyle,dua,chertkov,puliafito,celani,
gera,winkler,bhatt}. In biological systems, biomolecules subjected to
complex fluid flows~\cite{chu1,dua} are quite common, and a shear flow is
one such example. In shear flow, a polymer gets stretched as well as
tumbles in an irregular fashion. A crucial quantity which describes the
interesting conformational evolution of the polymer is its end-to-end
vector $\vect{R}$ (see Fig. 1). Recently, experiments on a single DNA
molecule in shear flow~\cite{gera} have obtained accurate probability
distribution functions of the length, the orientational angles, and the
tumbling times of the vector $\vect{R}$.  On the other hand, theoretically,
although scaling results from studies of non-linear single bead-spring
model~\cite{chertkov,celani,puliafito} and approximate analysis of
semi-flexible chains~\cite{winkler} are known, these are mostly non-exact.

For a non-linear system as a semi-flexible polymer (like DNA), approximate
theoretical results as in~\cite{chertkov,winkler} are perhaps as best as
one can get. They agree well with the static properties seen in
experiments~\cite{gera}. On the other hand, exact and analytically accurate
results are very desirable for at least the flexible polymer problem. In
particular, there exists no theory for the tumbling time statistics of the
vector $\vect{R}$; heuristic arguments given in~\cite{chertkov, winkler}
are simply inadequate, as will be evident from our analysis below.  In this
Letter, we derive exact and analytically accurate results for the static and
dynamic properties of $\vect{R}$ of a flexible Rouse chain~\cite{Rouse} in
shear flow.

The stochastic process of our concern, namely the end-to-end vector
$\vect{R}(t)$ of a linear polymer is a Gaussian random variable and its
dynamics is non-Markovian. The aspect of Gaussianity makes it quite easy to
write down the static ``joint'' probability density function (PDF) of the
Cartesian components of the vector $\vect{R}$, and consequently the joint
PDF of the length $R$, latitude angle $\theta$ and the azimuthal angles
$\phi$~\cite{puliafito,winkler}. The first non-triviality is to get the
PDFs of the individual polar co-ordinates, namely, $F(R)$, $U(\theta)$ and
$Q(\phi)$, in the stationary state. While $Q(\phi)$ was
known~\cite{puliafito,winkler}, in this Letter we derive $F(R)$ and
$U(\theta)$ exactly for a linear chain.

\begin{figure}
\includegraphics[width=3.375in]{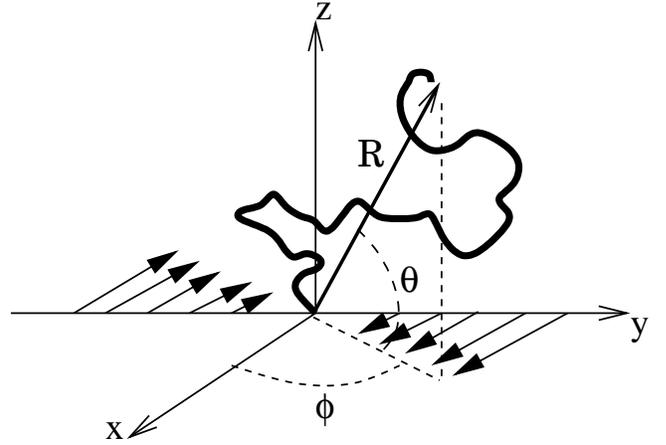}
\caption{A polymer configuration with shear flow along $x$-direction.}
\end{figure}

Secondly, the non-Markovian evolution of $\vect{R}(t)$ \cite{multi_Markov}
makes the first-passage questions (like the tumbling time statistics)
analytically extremely challenging. First passage questions in the context
of polymers have been of long standing interest
\cite{fixman,doi,marques}. In general, for non-Markovian processes,
calculation of first-passage properties are very nontrivial even when the
full knowledge of the non-exponential two-time correlation function is
available~\cite{McFadden,Review,satya,satya2}.  However, when a process is
{\it smooth} (as defined below) a method called `Independent interval
approximation' (IIA) is applicable~\cite{McFadden} and yields accurate
estimates~\cite{satya2,bhatt}. Very interestingly, while in the absence of
shear any component of $\vect{R}(t)$ is a non-smooth process and thus
analytical prediction is unknown, we show below that in the presence of
strong shear, a suitable component of $\vect{R}(t)$ associated with
tumbling becomes a smooth process, leading to analytical tractability via
IIA. Thus quite unexpectedly mathematical simplicity is achieved in a case
of greater physical complexity.  To be precise, we show that the PDF of
``angular tumbling times'' $\tau$, goes as $\sim \exp{(-{\alpha
\tau/\tau_0})}$ (where $\tau_0$ is the longest relaxation time of the
chain), and in the limit of large shear rate $\alpha\rightarrow 0.324^+$.
This number will serve as a lower bound for experiments.

As shown in Fig. 1, we study the Rouse dynamics~\cite{Rouse,edwards} of a
polymer chain of $N$ beads connected by harmonic springs, in a shear flow
in the $x$-direction. Let $\vect{r}_n(t)\equiv [x_n(t), y_n(t), z_n(t)]^T$
denote the coordinate vector of the $n^\text{th}$ bead ($n=1, 2,\dotsc,N)$
at time $t$. For $n=2,3,\dotsc,N-1$, the position vectors evolve with time
according to the equation of motion
\begin{equation}
  \frac{\rmd\vect{r}_n}{\rmd t} = k
  \left(\vect{r}_{n+1}+\vect{r}_{n-1}-2\,\vect{r}_n\right)+ \vect{f}_n(t) +
  \vect{\eta}(n,t),
  \label{evolve}
\end{equation}
where $k$ denotes the strength of the harmonic interaction between nearest
neighbor beads, and the vector $\vect{f}_n(t)\equiv [\dot{\gamma}
y_n(t),0,0]^T$ denotes the shear force field with rate $\dot{\gamma}$. The
Weissenberg number $\Wi = \dot{\gamma} \tau_0$, where the longest relaxation
time $\tau_0 = N^2/{k \pi^2}$. The vector
$\vect{\eta}(n,t)\equiv[\eta_1(n,t), \eta_2(n,t),\eta_3(n,t)]^T$
represents the thermal white noise with zero mean and a correlator
\begin{math}
\langle \eta_i(n,t) \eta_j(n',t')\rangle
= \zeta \delta_{ij}\delta_{n,n'} \delta(t-t'),
\end{math}
where $i,j=1,2,3$ and $n,n'=1,2,\dotsc,N$. The noise strength $\zeta$ is
proportional to the temperature and all the force strengths in
\eref{evolve} are scaled by viscosity. With free boundary condition, the
two end-beads (for $n=1$ and $n=N$) feel only one sided interaction and
therefore they evolve via modified equations which is obtained from
\eref{evolve} by using $\vect{r}_0(t)=\vect{r}_1(t)$ and
$\vect{r}_{N+1}(t)=\vect{r}_{N}(t)$, for two fictitious beads $0$ and $N+1$.

For large $N$ limit, the discrete $n$ of the beads is replaced by a
continuous variable $s$~\cite{edwards} and the discrete Laplacian in
\eref{evolve} is replaced by a continuous second derivative along $s$
direction. \Eref{evolve} then leads to
\begin{equation}
\frac{\partial \vect{r}(s,t)}{\partial t}
=k \frac{\partial^2 \vect{r}(s,t)}{\partial s^2} +\vect{f}(s,t)
+  \vect{\eta}(s,t),
\label{evolve continuum}
\end{equation}
with the free boundary conditions $\partial \vect{r}(s,t)/\partial s = 0$
at $s = 0$ and $s=N$.  In this continuum limit, the shear field and the
noise correlator are given by $\vect{f}(s,t)\equiv [\dot{\gamma}
y(s,t),0,0]^T$ and
\begin{math}
\langle \eta_i(s,t) \eta_j(s',t')\rangle = \zeta \delta_{ij}\delta(s-s')
\delta(t-t')
\end{math}
respectively. The end-to-end vector is $\vect{R}(t)
=\vect{r}(N,t)-\vect{r}(0,t)$.

Solving \eref{evolve continuum} by the Fourier cosine transformation
\begin{equation}
\vect{r}(s,t) =\tilde{\vect{r}}(0,t)+ \sum_{m=1}^{\infty}
\tilde{\vect{r}}(m,t) \cos\left(\frac{m\pi s}{N}\right),
\label{cosine series}
\end{equation} 
we find that the Fourier modes are given by
\begin{align}
\tilde{\vect{r}}(m,t) =& 
\int_{0}^{t} \Bigl[\tilde{\vect{f}}(m,t')
+ \tilde{\vect{\eta}}(m,t')\Bigr]\, 
\exp\bigl[-a_m (t-t')\bigr]\, \rmd t'\notag\\
&+\tilde{\vect{r}}(m,0)\, \exp(-a_m t)
\label{Fourier components}
\end{align} 
where $a_m = m^2/\tau_0$, and $\tilde{\vect{f}}(m,t)=[\dot{\gamma}
\tilde{y}(m,t),0,0]^T$ and
$\tilde{\vect{\eta}}(m,t)\equiv[\tilde{\eta_1}(m,t),
\tilde{\eta_2}(m,t),\tilde{\eta_3}(m,t)]^T$ is the $m^\text{th}$ Fourier
mode of the noise vector $\vect{\eta}(s,t)$.  The zero mode
\begin{math}
\tilde{\vect{r}}(0,t)=N^{-1}\int_0^N \vect{r}(s,t)\, \rmd s
\end{math}
describes the \emph{center of mass} motion of the polymer.
Equations~\eqref{cosine series} and \eqref{Fourier components}, together
with the knowledge of the Fourier space noise correlator
\begin{math}
\langle \tilde{\eta_i}(m,t) \tilde{\eta_j}(m',t')\rangle =2 \zeta N^{-1}
\delta_{ij}\delta_{m,m'} \delta(t-t'),
\end{math}
after some algebra, leads to the two point space and time dependent
correlation function between different relative position coordinates. In
the stationary state limit $t \rightarrow \infty$ with a finite time
increment $\tau\ge 0$, the correlation function
\begin{align}
  \bigl\langle[\vect{r}^{(i)}(s_1,t)
-\tilde{\vect{r}}^{(i)}(0,t)]\cdot[\vect{r}^{(j)}(s_2,t+\tau)-
  &\tilde{\vect{r}}^{(j)}(0,t+\tau)]
  \bigr\rangle   \notag\\
  \xrightarrow{t\rightarrow\infty} \frac{2\zeta}{N} \sum_{m=1}^{\infty}
  \cos\left(\frac{m\pi
      s_1}{N} \right) \cos\left(\frac{m\pi s_2}{N} \right) &\times  \notag \\
  \biggr[\delta_{i1} \delta_{j1} {\dot{\gamma}}^2 \left(\frac{2 \rme^{-a_m
        \tau}}{(2 a_m)^3} + \frac{\tau\rme^{-a_m \tau}}{(2 a_m)^2}\right) +
  &\delta_{ij}
  \frac{\rme^{-a_m \tau}}{(2 a_m)} \notag \\
  +\delta_{i1} \delta_{j2} \dot{\gamma} \left(\frac{\rme^{-a_m \tau}}{(2
      a_m)^2} + \frac{\tau\rme^{-a_m \tau}}{(2 a_m)}\right)& \biggr],
\label{corr}
\end{align} 
where ${i,j}={1,2,3}$ and the notations $\vect{r}^{(1)}(s,t)\equiv x(s,t)$,
$\vect{r}^{(2)}(s,t)\equiv y(s,t)$, and $\vect{r}^{(3)}(s,t)\equiv z(s,t)$.
Note that, any static correlation function in the stationary state limit
can be obtained by setting $\tau = 0$ in the above equation.  On the other
hand, for dynamic properties like tumbling, we need the correlators with
$\tau > 0$.

We first consider the static distribution functions related to
$\vect{R}\equiv[R_x, R_y, R_z]^T$. Since the Cartesian components are
Gaussian random variables, their joint PDF is,
\begin{math}
{\cal{P}}(R_x,R_y,R_z) =(2\pi)^{-3/2} |\vect{C}|^{-1/2}\exp(-\frac{1}{2}
\vect{R}^T \vect{C}^{-1}\vect{R}),
\end{math}
where $\vect{C}$ denotes the covariance matrix.  Putting $\tau = 0$ and
$(s_1,s_2) = (0,0), (0,N), (N,0)$ and $(N,N)$ in \eref{corr} we get the
four correlators, respectively, needed to calculate each elements of
$\vect{C}$. Since the vector $\vect{R}$ is a difference of two position
coordinates, the zero modes cancel, and hence we need not worry about it
being present in \eref{corr}. We find,
\begin{equation}
\vect{C}
\equiv
\left[\begin{array}{ccc}
 \langle R_x^2\rangle &  \langle R_x R_y \rangle & \langle R_x R_z \rangle \\
 \langle R_x R_y \rangle & \langle R_y^2  \rangle & \langle R_y R_z \rangle \\
 \langle R_x R_z \rangle & \langle R_y R_x \rangle & \langle R_z^2 \rangle 
\end{array}
\right] 
= \left[ \begin{array}{ccc}
 d & b & 0 \\
 b & a & 0 \\
 0 & 0 & a \end{array}
\right], 
\end{equation}
where
\begin{equation}
a = \frac{\zeta N}{2k}, \quad
b = \frac{\zeta N \pi^2}{48 k} \Wi , \quad
d = a + \frac{\zeta N \pi^4}{480 k} \Wi^2,
\end{equation}
and the determinant $|\vect{C}|=a c$, with $c = ad-b^2$.

The scaling of $\langle R_x^2 \rangle = d \approx N \Wi^2 \sim N^5$ is
a pathological feature of the Rouse model in shear flow in comparison
to semi-flexible polymers (better modeled with FENE constraint 
on $\vect{R}$ \cite{dua}). If the $\Wi$ dependencies are ignored, 
the asymptotic functional forms of the PDF's of the Rouse model, 
compare quite well with semi-flexible polymers.   

It is straightforward to obtain the joint PDF of the polar coordinates by
using the standard transformation: $\tilde{\cal P}(R,\theta,\phi)= {\cal P}
(R\cos\theta \cos\phi,R \cos\theta \sin\phi, R \sin\theta) R^2 \cos\theta$.
By integrating it over $R$ we get the joint angular PDF
\begin{align} 
S(\theta,\phi) = & \frac{|\vect{C}|\cos\theta}{4\pi} \bigl[ a \cos^2 \theta
\bigl\{ a \cos^2 \phi + d \sin^2 \phi \notag \\ - & 2b \sin \phi \cos \phi
\bigr\} + c \sin^2 \theta \bigr]^{-{3/2}}.
\label{joint} 
\end{align}
Now again integrating over $\phi$ in \eref{joint}, we obtain our first
important result, the PDF of the latitude angle,
\begin{equation}
U(\theta) = \frac{|\vect{C}|\cos{\theta}}{\pi (d_1 - d_2)\sqrt{d_1+d_2}}
\;\mathbf{E}\left(\sqrt{\frac{2 d_2}{d_1+d_2}}\right),
\label{Ttheta}
\end{equation} 
where $d_1 = [a (d+a)/2] \cos^2\theta + c \sin^2\theta$, $d_2 = (a/2)
[(d-a)^2 + 4 b^2]^{1/2} \cos^2\theta$, and
$\mathbf{E}(q)=\int_0^{\pi/2}\bigl(1-q^2\sin^2\beta\bigr)^{1/2} \rmd\beta$
is the complete elliptic integral of second kind~\cite{grad}.  From our
exact result in \eref{Ttheta}, by taking the limit of $\dot{\gamma} \gg 1$
and $\theta \ll 1$, we see that the $\mathbf{E}(.)  \rightarrow {\rm
constant}$, and $(d_1+d_2) \approx a d$ and $(d_1 -d_2) \approx c
\theta^2$: these lead to $U(\theta) \sim \Wi^{-1} \theta^{-2}$. Further,
from \eref{joint}, in the same limit $S(\phi=0,\theta) \sim \Wi^{-1}
\theta^{-3}$.  The azimuthal angle distribution
\begin{math}
Q(\phi) = \int_{-\pi/2}^{\pi/2} S(\theta,\phi)\,  \rmd\theta 
\end{math}
can be easily derived from \eref{joint}; we skip its explicit expression as
similar result has been derived earlier in \cite{puliafito,winkler}.  We
just note that $Q(\phi)$ peaks exactly at
\begin{math} 
\phi = \phi_m = \frac{1}{2} \tan^{-1} \left(\frac{2b}{d-a} \right).
\end{math} 
The full width at half maximum $\Delta \phi$ of $Q(\phi)$ is given exactly by 
\begin{math}
\cos(2\Delta \phi) = 2 - [(d+a)/(d-a)] \cos{2 \phi_m} 
\end{math}
and for $\dot{\gamma} \gg 1$, $\Delta\phi \approx \phi_m \sim \Wi^{-1}$
and \begin{math}
Q(\phi) \approx \Wi^{-1} \sin^{-2} \phi.  
\end{math} 
The asymptotic dependences of the functions $U(\theta)$, $S(\phi=0,\theta)$
and $Q(\phi)$ on $\theta$ and $\phi$, that we derive from our exact
Eqs. (\ref{joint}) and (\ref{Ttheta}) for a Rouse polymer, match with
earlier studies on semi-flexible
polymers~\cite{chertkov,puliafito,celani,winkler,gera}.

To derive our second result for the radial length distribution function
\begin{math}
F(R) = \int_{0}^{2\pi} \rmd\phi \int_{-\pi/2}^{\pi/2} \rmd\theta\,
\tilde{\cal P}(R,\theta,\phi),
\end{math} 
we employ the trick of first calculating the Laplace transform of the PDF of
$R^2$ instead, namely
\begin{math}
h(s) \equiv
\langle \exp(-sR^2)\rangle = \langle
\exp\bigl(-s[R_x^2 + R_y^2 + R_z^2]\bigr)\rangle.
\end{math} 
This is easily obtained as
\begin{math}
h(s)=\bigl(1 + 2as\bigr)^{-1/2}\,\bigl(1 + 2s(a + d) + 4 c s^2\bigr)^{-1/2}.
\end{math}
The PDF $F(R)$ is then related to the the inverse Laplace transform 
of $h(s)$ as 
\begin{math}
F(R) = 2 R\times \left[\mathcal{L}_s^{-1} \{ h(s)\} (R^2)\right]
\end{math}
and is given exactly as 
\begin{equation} 
F(R) = \frac{R^2 \rme^{-R^2/{2a}}}{\sqrt{2\pi |\vect{C}|}} \int_{0}^{1}
\rmd x \frac{\rme^{\lambda R^2 x} I_0(\mu R^2 x)}{\sqrt{1 - x}},
\label{FR3}
\end{equation}
where $I_0(.)$ is zeroth order modified Bessel function of first
kind~\cite{grad}, and
\begin{math}
\lambda =1/(2a) - (a+d)/(4c) 
\end{math}
and  
\begin{math}
\mu = \sqrt{(d-a)^2 + 4b^2}/(4c). 
\end{math}
From asymptotic analysis of \eref{FR3} we see that 
\begin{math}
F(R) \approx \sqrt{2} R^2\big/\sqrt{\pi |\vect{C}|}
\end{math}
for small $R$,
and 
\begin{math} 
F(R) \approx \exp\bigl(- [1/(2a) - \lambda - \mu] R^2\bigr)\big/
{\sqrt{4\pi |\vect{C}| \mu(\mu+\lambda)}}
\end{math}
for large $R$.

We now turn our attention to our third and main result on the first passage
question of the polymer ``tumbling'' process in shear flow. The tumbling
event is either defined as a radial return of the polymer to a coiled state
(as in experiments~\cite{gera} and simulations~\cite{celani}), or as a
angular return of vector $\vect{R}$ to a fixed plane~\cite{celani}, say
$\phi = 0$. The former radial definition relies on an arbitrary choice of a
threshold radius~\cite{gera, celani}, while the latter angular tumbling is
not. In this Letter we study the statistics of angular tumbling time,
i.e. the distribution of times $\tau$ between two successive zero crossings
of the stochastic process $R_x(t)$. For the scaled time $T = \tau/\tau_0$
the relevant PDF asymptotically is
\begin{math}
P(T) \sim \exp(-\alpha T). 
\end{math}
Analytical scaling dependence of $\alpha$ on $\Wi$ is known for
semi-flexible polymers \cite{chertkov}, but accurate constant factors were
not estimated.  We show below that for a Rouse chain, in the limit of large
$\Wi$, $\alpha$ approaches a constant value, and that can be estimated by
using a systematic IIA calculation.

For a Gaussian stationary process, the mean density of zero crossings is
given by~\cite{rice}: $\rho = 1/\langle T \rangle = \sqrt{-A''(0)}/\pi$,
where $A(T)$ is the normalized correlator, i.e., $A(0)=1$. We need
$A'(0)=0$ and a finite $A''(0)$ for $\rho$ to be finite ---then the process
is smooth and one can use IIA~\cite{McFadden}.

Now, for the relevant stochastic process $R_x(t)$ of our concern, using
\eref{corr} we find the stationary state correlator $C_{xx}(T) =
\lim_{t\rightarrow\infty}\,\langle R_x(t) R_x(t+\tau_0 T) \rangle$ as,
\begin{align}
C_{xx}(T) =& \frac{\zeta \tau_0 {\Wi^2}}{N} \sum_{m=1,3,5,\dotsc}^{\infty}
\biggl[\frac{\rme^{-m^2 T}}{m^6} + \frac{T\rme^{-m^2 T}}{m^4}\biggr]
\notag\\ &+ \frac{2\zeta \tau_0 }{N} \sum_{m=1,3,5,\dotsc}^{\infty}
\frac{\rme^{-m^2 T}}{m^2}.
\label{CT} 
\end{align}
The two sums in the first and the second lines in \eref{CT} for $C_{xx}(T)$
will be henceforth referred to as $C_\text{sh}(T)$ (due to shear) and
$C_\text{th}(T)$ (due to thermal fluctuations) respectively.

In the absence of any shear ($\Wi = 0$), the normalized correlator becomes
$A(T)= C_\text{th}(T)/C_\text{th}(0)$. Using $C_\text{th}(T)$ from
\eref{CT} we see that both $A'(0)$ and $A''(0)$ diverge, which in turn
makes $\rho$ infinite.  Thus, in this case $R_x(t)$ is non-smooth ---see
inset (b) of Fig. 2.  Although IIA fails in this rather simple looking
case, our numerical simulation gives $\alpha \simeq 1.20$ (see the curve
for $\Wi = 0$ in Fig. 2).

On the other hand, in shear flow ($\Wi \neq 0$), both the terms
$C_\text{sh}(T)$ and $C_\text{th}(T)$ are present in $C_{xx}(T)$, and the
small $T$ singular behavior of $C_\text{th}(T)$ contribute also to
$C_{xx}(T)$.  Thus although $R_x(t)$ has long excursions (see inset (a) of
Fig. 2) the thermal noisy contributions keep it non-smooth. While this
makes application of IIA seem hopeless, we note that for $\Wi^2 \gg 2$ in
\eref{CT}, the term $C_\text{th}$ can be ignored compared to $C_\text{sh}$.
More precisely, in the limit $\Wi \rightarrow \infty$, the normalized
correlator $A(T)\rightarrow C_\text{sh}(T)/C_\text{sh}(0)$.  Using
$C_\text{sh}(T)$ from \eref{CT}, one finds $A'(0)=0$ and
$A''(0)=-120/\pi^4$, giving a finite mean density of zero crossings $\rho =
\sqrt{120}/\pi^3$. The fact that the process $R_{x}(t)$ becomes smooth is
clearly seen in the inset (c) of Fig. 2. Thus in this limit of strong
shear, IIA becomes applicable.

A crude estimate of $\alpha$ can be made by approximating $P(T)$ to be
exponential over the full range of $T$ and not just asymptotically ---this
gives $\alpha \simeq \rho = 0.353$.  For a more systematic approach one
needs to use IIA. In Ref.~\cite{McFadden}, few different IIA schemes were
discussed.  We calculated $\alpha$ by all these various schemes, and the
various estimates differ slightly ---these details will appear in a future
publication. In this Letter, we present a particular approximation which
yields $\alpha$ very close to numerics.  We start with
\footnote{This approximation, which is an exact equality for the correlator
of the clipped process $\text{sgn}(R_x)$, is used here for analytical
tractability. Moreover, it yields a closer match to the numerics.}
$A(T)\approx \sum_{n=0}^\infty (-1)^n p_n(T)$, where $p_n(T)$ is the
probability of having $n$ zero crossings of $R_x$ between $0$ and $T$. Then
IIA assumes $p_n(T)$ to be a product of the probabilities of intervals
which make up the stretch $0$ to $T$, integrated over the locations of the
zero crossings. The latter convolution integrals are best handled by
Laplace transformation, and one eventually obtains a relation between the
Laplace transforms ${\tilde A}(s)$ and ${\tilde P}(s)$, of $A(T)$ and
$P(T)$, respectively~\cite{McFadden,satya2}:
\begin{math}
  {\tilde P}(s) = \bigl[{1 - (\langle T \rangle/2)s(1-s{\tilde
  A}(s)}\bigr]\big/ \bigl[{1 + (\langle T \rangle/2)s(1-s{\tilde
  A}(s)}\bigr].
\end{math}
From \eref{CT}, we obtain the exact Laplace transform of $C_\text{sh}(T)$
and hence ${\tilde A}(s)$ in the limit $\Wi\rightarrow\infty$ as,
 \begin{align}
 {\tilde A}(s) = \frac{1}{s} - \frac{120}{\pi^4 s^3}  &- \frac{60
  }{\pi^4 s^3}\, \mathrm{sech}^2\left(\frac{\pi \sqrt{s}}{2}\right) \notag\\
 &+  \frac{360}{\pi^5 s^{7/2}}\,
   \mathrm{tanh}\left(\frac{\pi \sqrt{s}}{2}\right). 
 \end{align}
Since $P(T) \sim \exp(-\alpha T)$, the Laplace transform ${\tilde P}(s)$
must have a simple pole at $s = -\alpha$. In other words, the denominator
of ${\tilde P}(-\alpha)$ must vanish, i.e.  $1 - (\langle T
\rangle/2)\alpha(1 + \alpha{\tilde A}(-\alpha)) = 0$, where $\langle T
\rangle = 1/\rho = \pi^3/\sqrt{120}$. Solving for $\alpha$ from the latter,
we finally have,
\begin{equation}
\alpha = 0.323558.... 
\label{cIIA}
\end{equation}

\begin{figure}
\includegraphics[width=3.375in]{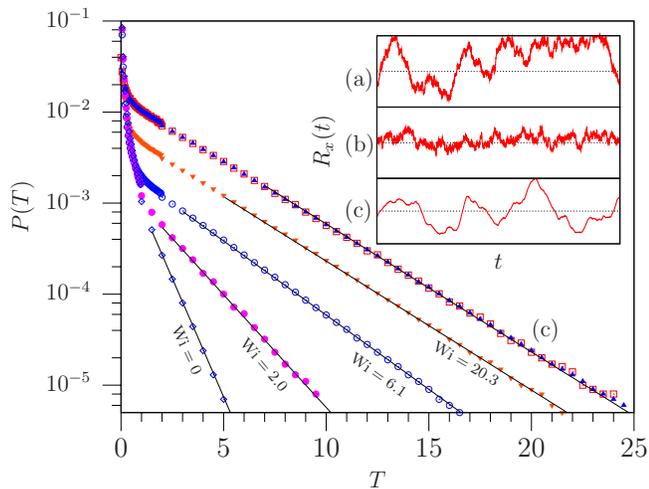}
\caption{Linear-log plot of $P(T)$ versus $T$: the fitted $\alpha$'s for
  the four curves with $\Wi=0, 2.0, 6.1$, and $20.3$ are $1.20, 0.57,
  0.375$, and $0.326$ respectively.  The two datasets with symbols
  {\color{red} $\boxdot$} and {\color{blue}$\blacktriangle$} in (c) are
  obtained by switching off the thermal noise along $x$ direction
  ($\eta_1=0$) in \eref{evolve} and $\dot{\gamma} = 0.2$ and $0.6$
  respectively ---both fit well with the analytical $\alpha = 0.324$
  line. All data are for $N = 10$.  Inset: Typical $R_x(t)$ versus $t$
  corresponding to the cases (a) both $\dot{\gamma}\not=0$ and
  $\eta_1\not=0$, (b) $\dot{\gamma}=0$ and $\eta_1\not=0$, and (c)
  $\dot{\gamma}\not=0$ and $\eta_1=0$.}
\end{figure}

To check the accuracy of our analytical result \eref{cIIA}, we perform a
simulation switching off the thermal noise in the $x$ direction
($\eta_1=0$) in \eref{evolve} ---this effectively achieves the limit $\Wi
\rightarrow \infty$ for any finite $\dot{\gamma}$. For the latter case,
with $\dot{\gamma} = 0.2$ and $0.6$, we show in Fig. 2 that their slopes
$\alpha$ for $P(T)$ have excellent agreement with \eref{cIIA}. For any
finite $\Wi$ (keeping $\eta_1\not=0$), the value of $\alpha$ smoothly
interpolates between the two limits $\simeq 1.20$ and $0.324$ (see Fig. 2).

No direct comparison can be made with the published experimental
data~\cite{gera}, as the latter study is for radial tumbling.  We look
forward to future experiments on angular tumbling of a semi-flexible
polymer.  We claim that our result for $\alpha$ in \eref{cIIA} will serve
as a lower bound, based on the following argument. For the small $\Wi$
regime, a semi-flexible polymer may be represented by the Rouse limit, for
which we have shown (Fig. 2) that $\alpha$ decreases as $\Wi$ increases and
approaches the value in \eref{cIIA} from above. On the other hand, for the
large $\Wi$ regime, it is known from experiments \cite{gera} and FENE model
simulations \cite{celani} that $\alpha$ increases as $\Wi$ increases.
These two facts put together imply that $\alpha$ would reach a minimum
value for some intermediate $\Wi$ and that can only approach the value in
\eref{cIIA} from above.  In summary, we have obtained exact PDFs for the
length and latitude angle of the end-to-end vector of a Rouse polymer in
shear flow. Further, we have derived an accurate analytical estimate of the
decay constant associated with the PDF of angular tumbling times for a
Rouse chain in the limit of strong shear.

\begin{acknowledgments}
We thank S.N. Majumdar and A. Sain for useful discussions, and grant
no. $3404-2$ of ``Indo-French Center for the Promotion of advanced research
(IFCPAR)/CEFIPRA)''.
\end{acknowledgments}

\end{document}